\documentclass[journal]{IEEEtran}
\usepackage{cite}

\ifCLASSINFOpdf
  \usepackage[pdftex]{graphicx}
\else
\fi
\usepackage{amsmath}
\usepackage{algorithmic}

\usepackage{array}

\usepackage{stfloats}
\usepackage{url}

\begin{document}
%

\title{The Responsible Development of\\Automated Student Feedback with Generative AI}

%
%
%


\author{Euan~D~Lindsay,~\IEEEmembership{Senior Member,~IEEE,}
        Mike~Zhang,
        Aditya~Johri,
        and~Johannes~Bjerva
\thanks{E.\ D.\ Lindsay is with the UNESCO Centre for Problem Based Learning at
Aalborg University, Thomas Manns Vej 25, 9220 Aalborg Ø, Denmark. Email: \texttt{edl@plan.aau.dk}}
\thanks{M. Zhang is with the Department of Computer Science, Aalborg University,
A.C. Meyers Vænge 15, 2450 København SV, Denmark. Email:
\texttt{jjz@cs.aau.dk}}%
\thanks{A. Johri is the Director of the Technocritical Research on AI, Learning \&
Society Lab (trailsLAB) at George Mason University, Fairfax, VA 22030.
Email: \texttt{johri@gmu.edu}}
\thanks{J. Bjerva is with the Department of Computer Science, Aalborg University,
A.C. Meyers Vænge 15, 2450 København SV, Denmark. Email:
\texttt{jbjerva@cs.aau.dk}}%
}

\maketitle
\begin{abstract}

Providing rich, constructive feedback to students is essential for supporting and enhancing their learning. Recent advancements in Generative Artificial Intelligence (AI), particularly with large language models (LLMs), present new opportunities to deliver scalable, repeatable, and instant feedback, effectively making abundant a resource that has historically been scarce and costly. From a technical perspective, this approach is now feasible due to breakthroughs in AI and Natural Language Processing (NLP). While the potential educational benefits are compelling, implementing these technologies also introduces a host of ethical considerations that must be thoughtfully addressed.

One of the core advantages of AI systems is their ability to automate routine and mundane tasks, potentially freeing up human educators for more nuanced work. However, the ease of automation risks a ``tyranny of the majority'', where the diverse needs of minority or unique learners are overlooked, as they may be harder to systematize and less straightforward to accommodate. Ensuring inclusivity and equity in AI-generated feedback, therefore, becomes a critical aspect of responsible AI implementation in education.

The process of developing machine learning models that produce valuable, personalized, and authentic feedback also requires significant input from human domain experts. Decisions around whose expertise is incorporated, how it is captured, and when it is applied have profound implications for the relevance and quality of the resulting feedback. Additionally, the maintenance and continuous refinement of these models are necessary to adapt feedback to evolving contextual, theoretical, and student-related factors. Without ongoing adaptation, feedback risks becoming obsolete or mismatched with the current needs of diverse student populations. Addressing these challenges is essential not only for ethical integrity but also for building the operational trust needed to integrate AI-driven systems as valuable tools in contemporary education. Thoughtful planning and deliberate choices are needed to ensure that these solutions truly benefit all students, allowing AI to support an inclusive and dynamic learning environment.



\end{abstract}


%
\IEEEpeerreviewmaketitle

\section{Introduction}
\begin{figure}[!t]
\centering
\includegraphics[width=\columnwidth]{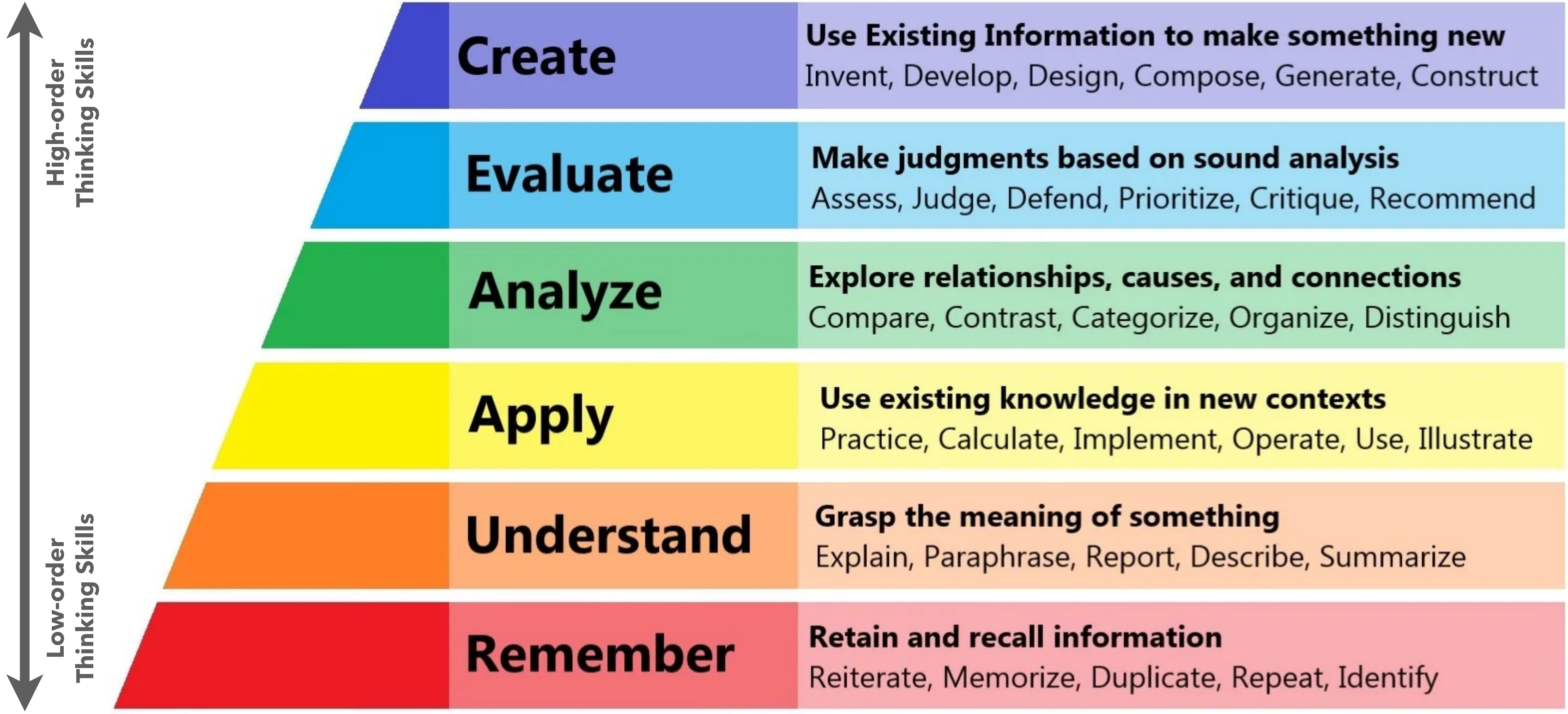}
\caption{A visualisation of Bloom's revised taxonomy, modified from~\cite{bloom-taxonomy}.}
\label{fig:bloom}
\end{figure}

%
%
%
%
The release of powerful language technology tools based on generative language modelling (e.g., ChatGPT, GPT-4(o), Claude, Gemini, Llama; \cite{nikolic2023chatgpt, geminiteam2024gemini, touvron2023llama}), marked a significant shift in how we approach higher education. 
For example, days after the release of ChatGPT, students, educators, and the public alike discovered the potential of the application for assisting with a range of teaching and learning tasks, but also encountered significant challenges in terms of academic integrity.
The response of many leading technical universities was to revert to pen-and-paper formats in exams~\cite{cassidy2023}. 
Previous work, e.g., \cite{nikolic2023chatgpt}, find that ``with little modification to the input prompts, ChatGPT could generate passable responses to many of the assessments''. 
The intense focus on academic integrity triggered an important and necessary conversation about the role of assessments, in particular, the increased potential of automatic assessment. 
Where prior approaches to automated assessment have been limited to, e.g., multiple-choice style questions, generative language modelling completely removes this barrier, potentially allowing for assessment of any type of student output. 
The need for developments in this space is clear given the recent developments, as access to tools such as ChatGPT potentially obliterates assessment at lower levels of Bloom’s hierarchy~\cite{Bloom1956} (Fig.~\ref{fig:bloom}), and it is an open question how writing with generative language models affects writing processes and assessment thereof. 
After all, if professionals are going to use AI tools in their working lives, we should aim to train them in their use.

While assessment is a clear space of development for this type of educational technology, we argue that the real potential of generative language modelling can be found in student feedback. 
We propose that this type of technology can, relatively straightforwardly, be developed such that a student in any educational program can essentially receive unbounded amounts of feedback. 
One advantage of this approach over current approaches is that the existing feedback mechanisms using, e.g., multiple choice questions or parametrized questions in engineering courses are limited by several factors.
For one, such methods do not easily scale, and tend to tie educational programs into a strict set of practices, where it is difficult to argue for innovation due to the prohibitive costs of developing new feedback scripts. 
Although many advantages of automation are gained from these practices, e.g., recognize common patterns of student answers and standardize responses to them, rather than having to make bespoke responses to them all, there is a further weakness in that this feedback does not extend to a large portion of students. 
\emph{We argue that current approaches to automated feedback best serve the median student, as developing tailored feedback for the long tail of students---be it highly excelling students who can be helped to excel further, or struggling students who need specific help to succeed in their education---is more expensive, both in terms of human and fiscal resources, and helps fewer students}. 
The flexibility and scalability of generative language modelling provide a unique solution to this tyranny of the long tail(s). 
This, in turn, allows us to consider real universal coverage using automated feedback, allowing us to meet our obligations to all students, not just those who present the most common submissions. Although numerous successful case studies (e.g., \cite{urban2024chatgpt}) highlight ChatGPT's potential for fostering, e.g., creative problem-solving, the ethical implications of integrating such systems into classroom settings have yet to be thoroughly examined.

To move towards this goal, we must consider how we generate this feedback, who engages with this feedback (and when and how), and the impact of providing feedback in a new paradigm. 
This requires the development of a framework for responsible development, where we follow recommendations from previous work in that we look at the dangers of automation in engineering education, aiming to keep engineering education human-centered \cite{kasneci2023chatgpt,wang2024large}. 
While doing so we must avoid the ``Turing Trap'' (i.e., realistic human intelligence automation via an excessive focus on automation based on AI technologies)~\cite{brynjolfsson2022turing}. In Fig.~\ref{fig:framework}, we visualize our contributions. Overall, the development of this framework denotes the core contribution of this article, in which we:

\begin{figure*}
    \centering
    \includegraphics[width=.8\linewidth]{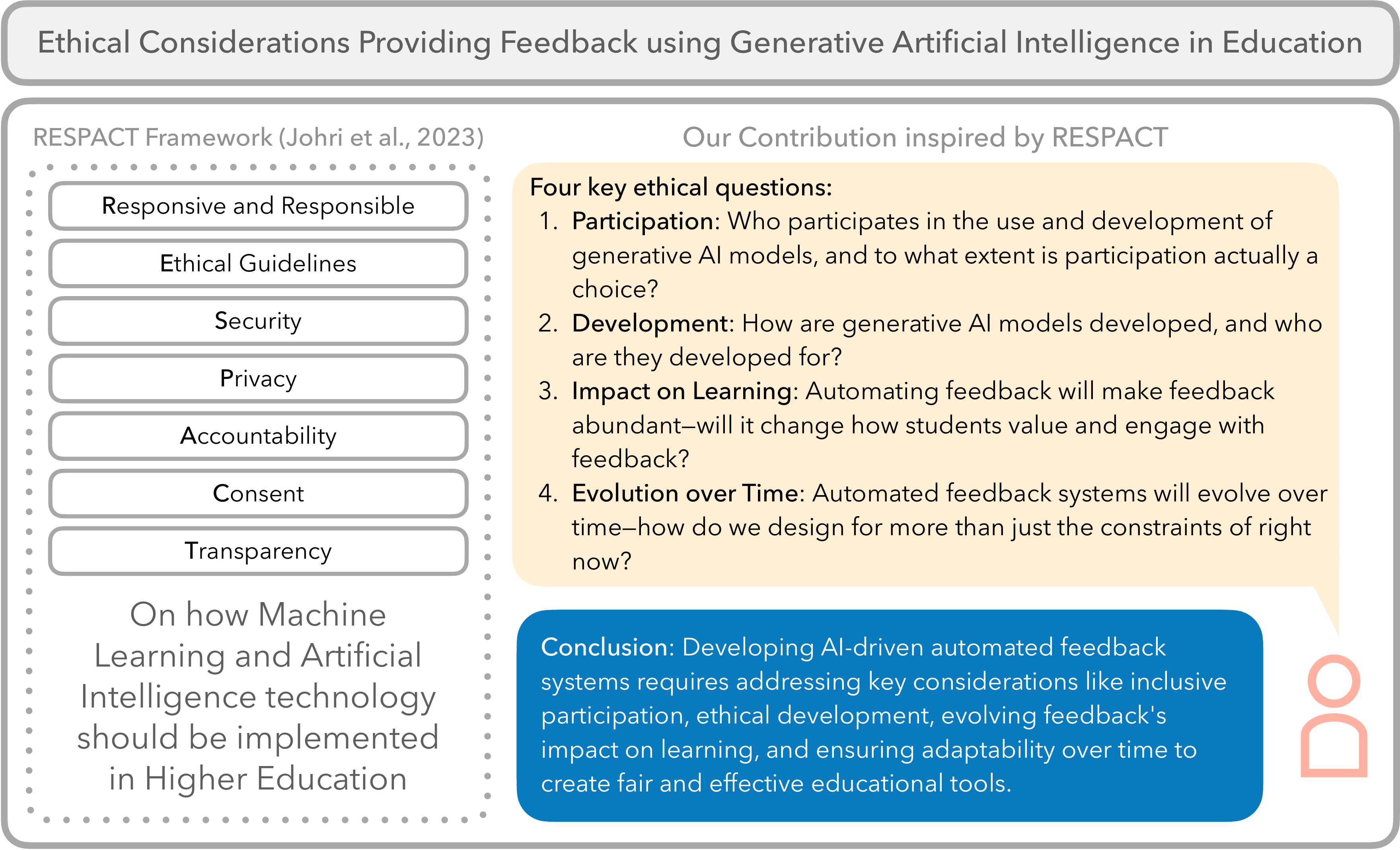}
    \caption{We explore critical ethical considerations for using generative AI to deliver automated student feedback in education. By applying an ethical framework for AI and machine learning~\cite{johri2023ethical, johri2025ethical}, this study addresses both development and maintenance phases, emphasizing how these tools will evolve over time. Key ethical issues highlighted include inclusive participation, responsible development, impact on student learning, and adaptability over time.}
    \label{fig:framework}
\end{figure*}

\begin{itemize}
    \item Outline the current frontiers of automated feedback;
    \item Apply the RESPACT framework to guide academics in responsibly developing automated feedback systems, with a focus on \textit{technology}, \textit{computing}, and \textit{engineering}-related teaching domains;
    \item Identify the ethical issues involved in the provision of automated feedback.
\end{itemize}

\section{The Frontiers of Automated Feedback}
Engineering education has a long history of automating assessment, primarily focused on reducing the workload of assessing the lower-order thinking at the lower levels of Bloom's taxonomy (Fig.~\ref{fig:bloom}).  Multiple choice questions can test memory and understanding but require simplified questions. Parameterized numerical questions can test application and analysis of engineering theory, while offering the possibility of providing individualized versions of questions to each student. Automated test suites can compile, validate and test student code; but both require significant initial investment in anticipating student responses, and can only provide feedback if the students make anticipated errors. Adaptive learning systems allowed for identification of common student mistakes, and for providing tailored feedback given those mistakes. However, the workload needed to develop comprehensive feedback for any new learning material is prohibitive, and only ``pays off'' given many students, and/or static educational material across time. 

This also entails the converse, namely that given an early investment in this type of learning materials, new developments are difficult to invest in. Hence, this current state has effectively locked some engineering courses into a focus, where a particular set of questions are iterated over. If a faculty member has put significant effort into identifying (say) the five most common errors, and the 2$\textsuperscript{5} =$ 32 possible combinations of students making all, some or none of them, and then set up the parameterized equations to identify each of those different combinations and provide the right set of feedback for each combination, there will be an inevitable reluctance to throw them away and build a new question from scratch next year. The issue is precisely that the current practice has not been to build a system to give feedback to students---but rather to build a system to give feedback on a \emph{single question} to students. This clearly does not scale without significant resource consequences.

Looking at the top of Bloom's pyramid (Fig.~\ref{fig:bloom}), there is a near complete gap in engineering education literature regarding automation of assessment and feedback. Prior to the release of generative tools such as LLMs, automation of the higher levels (evaluating, creating) was seen as unfeasible, as this type of assessment was out of reach of the existing AI-tools. Existing models can provide adaptive feedback, but that feedback is developed in advance, and deployed to the students based on guidelines, rules, and metrics that are developed in advance. This has proven to be a successful manner to deal with the center mass of the distribution of students. For instance, if 40\% of the wrong answers are a specific misconception (e.g., using static friction when you should use dynamic friction) then there is a huge leverage to be had in pre-preparing an answer and feeding it back to the students automatically. 

One core challenge in previous work is that we want to be able to deal with the long tail of required feedback, where some feedback is clearly necessary, but falls outside of the scope of what can feasibly be prepared in advance. The current automated feedback systems do not tailor feedback to these, but tend to default to a default response. An LLM, without further development, would at least return something, but would include the risk of providing a ``hallucination'' unrelated to the learning goals or problem at hand. Can frontier feedback models allow us to update our questions each year, or provide new scenarios? Can this type of technology keep up with a slow drift in the kinds of answers, reflections, or insights that students need?  As curricula move from what students should learn, to what graduates should be able to do, to who professionals should be, we get more nuanced and subtle changes and evolutions in what we are actually looking for from our students, and so we possibly need a more continuous evolution in the feedback we give them. In essence, this entails a transition from compile time flexibility to run time flexibility.

\subsection{What do Lange Language Models Offer?}
Since the release of, e.g., ChatGPT, experts in education technology have started exploring the potential of this type of tool in automating processes in education. Assessment has been at the forefront of the field's attention~\cite{nikolic2023chatgpt}---while assessment clearly is a use case, the possibilities offered by generative language modelling are not limited to this relatively basic application. Instead, we here focus on what we argue is a more viable and impactful long-term implementation of language technology in education, namely automation of feedback. 

In short, we argue that LLMs offer opportunities to provide automated feedback, at the top of the pyramid. The technical developments needed to bring generative language modelling to this step are fairly foreseeable. What is more difficult, however, is anticipating how this might impact learning, and how this can be implemented in a responsible manner from an ethical perspective. It goes without saying that this type of tool will inevitably change the paradigm of feedback, as this will make feedback an abundant resource for students.

\subsection{The Potential of Generative Language Modelling}
Before looking into ethical perspectives, we first outline the technical details of LLMs in its current state, leading into what technological developments are needed to develop a generative language model-based system for abundant feedback in learning situations.

Models such as ChatGPT are typically referred to as a dialogue-based chatbot, and likewise the growing interest around generative AI seems to be centered around this type of chatbot-esque interface. However, the real underlying technology is known as language modelling, and it is in clever applications of this type of model that the real potential lies. This terminology and technology stems from the field of Natural Language Processing (NLP) and deals with an interpretation of natural languages as strings of symbols, traditionally decomposed into “words”. A language model, then, simply has the goal of computing the probability of a sequence of words, e.g.,:

\begin{equation}\label{eq1}
    \text{P}(\text{W}) = \text{P}(w_1, w_2, ..., w_n)
\end{equation}

Where $w_n$ denotes the $n\textsuperscript{th}$ word in a sequence. This tends to be intractable with simple probabilistic approaches, and is typically reformulated into a conditional probability:

\begin{equation}\label{eq2}
    \text{P}(\text{W}) = \text{P}(w_n | w_1, w_2, ..., w_{n-1})
\end{equation}

These formulations are both equivalently known as a language model (LM). It is worthwhile to note that the choice of terminology here is debatable, as this is not exactly a model of language, given that language is not a strictly linear artefact. Nonetheless, this formulation turns out to be an effective tool, and has yielded virtually all recent advances known to the general public, as all modern approaches to NLP rely on LMs. Rather than simply calculating conditional probabilities by, e.g., counting frequencies of combinations of words known as n-grams, the past 5--10 years have yielded relatively sophisticated neural machine learning models, culminating in the transformer-based LMs of today, such as the GPT-models, Llama, Bloom, and others~\cite{touvron2023llama,le2023bloom}. The utility of a language model is not simply found in its ability to assign a probability to a sequence, although it does follow from this fact. For one, it is relatively straightforward to take the step from the conditional probability of a sentence in Eq.~\ref{eq2}, to a generative model which could, e.g., sample the most probable continuation of a sentence. Always sampling the most probable continuation would, however, not lead to a particularly natural output, hence effective and optimal sampling in LMs is a leading research area in NLP~\cite{holtzman2019curious,meister2023locally}.

\subsection{Automatic Feedback for Students in Natural Language Processing}
Feedback is a powerful means of supporting learning~\cite{hattie2008visible} to inform the learner about their actual state of performance~\cite{narciss2008feedback}. There is a wide array of previous work on automated feedback in NLP, even before LLMs~\cite{keuning2018systematic} in the context of feedback on programming exercises~\cite{jeuring2022towards, kiesler2023exploring, hellas2023exploring, jacobs2024evaluating, phung2024automating}. 

Additionally, there is other work in the context of student learning, peer-to-peer learning, or peer reviewing. They are focused on comprehensive overviews of peer-to-peer feedback~\cite{bauer2023using}, using sentence similarity methodologies for aligning answers to open responses in mathematical questions~\cite{botelho2023leveraging}, using language models to assist students improve their critical thinking skills by argumentation~\cite{guerraoui2023teach}, benchmarking whether GPT-style models~\cite{brown2020language,achiam2023gpt} align with feedback on research papers, and last, using LLMs in-the-loop during student programming assignments~\cite{liang2023can, pankiewicz2023large}. A current gap in the research for automatic feedback with LLMs is to evaluate the given generated output. Most work focuses on using additional LLMs-as-evaluator, i.e., scoring the generated output of one LLM by another one~\cite{han2023fabric, zheng2024judging, bai2024benchmarking, verga2024replacing}.

\subsection{How do we Build Appropriate Language Models?}
Here discuss the core technology of language modelling.  A good LM is an LM which produces a high probability for sentences which seem natural, and vice-versa. The key to constructing a good LM, beyond incremental improvements to LM architectures, is as in most parts of machine learning and AI; we need a massive amount of training data for downstream performance metrics. For an LM, the training data is typically a corpus, containing a large collection of unlabeled text. As an example, GPT-4 was trained on approximately 500 billion words~\cite{achiam2023gpt}.  Historically, language models have been built without much consideration for the quality or type of this data, other than considering the language at hand (English in most cases). Due to their substantial data requirements, modern LMs, are typically trained with even less consideration for the input data. When the need for training data is in the ballpark of hundreds of billions of words, one typically does not have the luxury of being too picky, hence in essence using all data within reach. This has the side-effect of offering multilingual capabilities, as unavoidably a mix of the languages on the internet are used in training. In sum, modern language models can be seen as a distillation of a large sample of the internet, with both positive and negative ramifications. This training paradigm comes with ethical considerations of its own, e.g., in the area of perpetuating cultural biases~\cite{cao2023assessing, navigli2023biases, kirk2024benefits, gallegos2024bias}, and amplifying extremist content online. In the long run, when more and more online content is generated by language models rather than humans, there will also be an issue with the AI echo-chamber, where language models trained on their own output will spiral into regurgitating more and more of this biased content. Furthermore, the underlying model has issues in terms of hallucinations, and a general lack of factuality~\cite{ji2023survey}. 

In addition to being based on an LM, a chatbot such as ChatGPT is further trained using Reinforcement Learning with Human Feedback (RLHF;~\cite{christiano2017deep, stiennon2020learning, ouyang2022training}). The technical details of this training are omitted here, but in essence this training allows users to have relatively natural typed conversations with the system.  The result is a general-purpose tool, which clearly shows potential, but lacks suitability to provide real value for learners. All-in-all, we have the tools needed to develop general-purpose chatbots, bringing us to the question: What do we need to add in order to develop specific-purpose technical solutions to delivering, e.g., unbounded feedback in learning situations?

\subsection{What Technical Innovations are Needed for Large Language Models to Provide Feedback?}
Given the current state of LLMs, and given the relative ease with which RLHF can be adapted to different applicational scenarios, there is a clear technical path forward towards providing abundant feedback. Let us consider the case of a specific course in, e.g., Algorithms and Data Structures in a computer engineering education. Given the versatility of LMs, a relatively straight-forward implementation could consist of further training the LM on course-specific literature as well as continued RLHF tuning with domain experts and a periodic system evaluation. Here, we see the potential to allow potential personalisation of these dialogue-based systems. There are several works on personalising LLMs for the user (e.g.,~\cite{liu2024personality}), including ethical considerations thereof~\cite{kirk2024benefits}.

This straightforward procedure, albeit costly in terms of computational and human resources, would likely yield an acceptable model in terms of user experience. One drawback of this paradigm is that we would likely need to repeat the steps for any new course or degree, as different outputs would be expected in different learning situations. In other words, given the same input, a feedback model should provide different outputs depending on factors such as the learning goals of the student.

This solution still has drawbacks, however. For instance,  the output is not verified, and it is difficult to expect all output to be sound feedback. This type of model would still suffer significantly from drawbacks in terms of factuality and hallucinations, and there would be no guarantees that any feedback given would actually be useful. Much like with the general-purpose LLM, output is designed to look natural, but its true utility is difficult to assess. More importantly, a training paradigm as outlined above would most likely be effective at providing feedback for our median students, but less likely to be able to support students in the long tail.

\subsection{Future Potentials in Learning Outcomes}
How does access to unlimited super-human feedback affect learning? We have a hypothesis or best-case scenario that using generative language modeling will allow us to push students the higher levels of Bloom's taxonomy, resulting in ``better learning'' and ``better students''. There is precedent for this assumption, when looking at other domains. Within the domain of board games, access to superhuman AI has drastically changed the field. All professional chess players use supercomputers in their preparational work and in analysis of games, and even amateurs frequently make use of such tools. A superhuman level in Go was reached in 2013 when AlphaGo surpassed the level of the reigning world champion. Interestingly,~\cite{shin2023superhuman} find that access to superhuman AIs drastically changed the playing field among professional Go players, based on data from the 20th and 21st centuries. Concretely, they find that unlimited feedback from superhuman AI significantly improved upon human decision-making processes, and allowed for humans to learn to play more novel moves than before training with this type of feedback. Can we expect the same type of outcome in higher education, if we can properly apply generative LMs for unlimited feedback?

Due to these recent developments in generative language modeling, and the clear potential of this type of methodology in educational technology, we foresee a significant ramp up in how educational institutes across the world make use of the technology. As we now have the tools at hand to provide potentially unlimited feedback to students, it may be tempting to jump straight into providing this feedback without second thought. However, rather than being preoccupied with whether we can provide this feedback and implementing it everywhere, we here stop to think whether we should, and how this can be done in a responsible manner.

\section{The Ethical Issues Involved in the Provision of Automated Feedback}
The provision of automated feedback is in the realms of technical feasibility, and tools such as ChatGPT are pushing the frontiers of what feedback can be provided adaptively at run-time rather than anticipated at compile time.  Providing such feedback inherently raises a range of ethical issues and responsibilities towards students. Some of these are obvious while others are subtle; some are inherent in the nature of automated feedback, while others represent parameters over which we have significant control. 

We derive our ethical considerations surrounding automated feedback from the RESPACT framework~\cite{johri2023ethical}. The framework is designed for AI and machine learning in higher education and includes \textbf{seven key dimensions}:

\begin{enumerate}
    \item \textbf{Responsive and Responsible}: Responsible use of generative AI involves ensuring feedback aligns with course learning goals. For generic models, feedback must be contextually relevant; for custom models, training must be rigorous and unbiased. Appropriateness involves both relevance and factuality, although current models lack mechanisms for ensuring factual accuracy.
    
    AI tools should also address common and uncommon student inputs, providing equitable support. All faculty must monitor how students use the tool and ensure it supports their learning goals. Allowing access to the tool for all students requires reliable digital connectivity and potentially third-party licensing.

    \item \textbf{Ethical Guidelines}: The RESPACT framework emphasizes the need for explicit ethical guidelines in implementing new technologies. Emerging technologies lack established usage examples, necessitating strong dependence on principles and guidelines. Explicit guidelines on the use of generative AI may help resolve ethical tensions and build trust by making institutional priorities clear.
    
    \item \textbf{Security}: Data security is crucial for educational technology. Student work is confidential, and the rise of authentic assessments increases the likelihood of sensitive data. Many automated feedback systems rely on external providers (e.g., ChatGPT), raising concerns about data ownership and security beyond mere server protection.
    
    \item \textbf{Privacy}: Automating feedback complicates privacy issues. Student assignments, now AI system inputs, must be protected as confidential information. Tailored feedback poses higher privacy risks. Evaluating AI systems involves additional privacy risks, particularly for unique student submissions.
    
    \item \textbf{Accountability}: Accountability frameworks help identify and record compromises in AI use. Defining success for automated feedback clarifies goals, measures impacts on different user groups, and ensures balanced cost-benefit distribution. Continuous review and monitoring build trust and allow for adjustments based on actual impacts.
    
    \item \textbf{Consent}: Generative AI's operational context complicates consent. Differentiated consent allows students to control specific data uses, addressing the all-or-nothing consent issue. Consideration must be given to third parties involved in student assignments. Students can opt out of using AI tools, but widespread adoption might limit their educational opportunities. For custom-trained models, consent acquisition should be straightforward, but outsourced models require thorough investigation of consent practices.

    \item \textbf{Transparency}: Transparency is essential for trust in generative AI systems. Explainable AI aims to make AI operations understandable. Transparency in model training builds trust in feedback. Open processes foster trust among students, faculty, accreditation bodies, employers, and government ministries, ensuring buy-in for AI systems that enhance student learning.
\end{enumerate}

These dimensions guide the application of generative AI for automated feedback, ensuring all students benefit. By these means, we propose the following \textbf{four ethical questions} with elaboration surrounding the provision of automated feedback and should be taken into consideration for developing these systems in the future. They fall into six key categories:
\begin{enumerate}
    \item \textbf{Participation}:  Who participates in the use and development of generative AI models, and to what extent is participation actually a choice?
    \item \textbf{Development}: How are generative AI models developed, and who are they developed for?
    \item \textbf{Impact on Learning}: Automating feedback will make feedback abundant---will it change how students value and engage with feedback?
    \item \textbf{Evolution over Time}: Automated feedback systems will evolve over time---how do we design for more than just the constraints of right now?
    \item \textbf{Scaling Up}: Automated feedback systems offer the opportunity of working at scale---how do we account for the broader scope?
    \item \textbf{AI Literacy}: To what extent will we rely upon the skills of the users of these systems?
\end{enumerate}

Several of these issues are ethical issues in AI more broadly; but they have specific manifestations in the automated feedback context. We elaborate on each point in the following sections.

\subsection{Participation}
A key part of the value proposition of automated feedback is its scalability---the ability to support large numbers of students without having to incur the marginal costs of the people to do so. Implicit in this model is that all of the students will participate---often without a meaningful choice to do so.

It will be impractical to seek consent from individual students as large-scale implementations are rolled out; operating a parallel non-automated feedback stream for students who do not consent will be far too costly, and will undo many of the organizational benefits of the process.  Even if a consent process is operated, there are power differentials between individual students and the academics and the institutions that will cloud the ability of individuals to freely choose to participate. Institutions will seek compulsory participation, partly because that simplifies the implementation, but further because automated feedback could be helpful for learning.  However, the students that can potentially benefit the most from such systems---those in the long tail---are also the ones who carry the greatest risks from participating, and who are most disempowered by removing the choice to opt out.

The issue of forced participation is even more challenging when the development of a corpus of training data is considered.  Universities inherently capture significant amounts of data about their students as they conduct their teaching operations, and share this with their learning management system providers; this is an accepted necessity of the operation of a digital learning environment.  Training sets for automated feedback are usually convenience samples of previous students’ work---data that were collected for the purpose of learning and assessment, not the development of new tools for the future benefit of others.  Larger, multi-year datasets add the additional complexity that it may no longer be possible to contact the students in the dataset as students graduate, leave, and move on to other (non-university) contact details.

The European General Data Projection Regulation (GDPR) has responsibilities regarding data re-use, and the minimization of the quantities and nature of data being collected.  A larger dataset will provide more information for an AI system to learn from; but this data is being taken from its original purpose and context, usually without the consent nor knowledge of the people those data represent.  Issues with data privacy are exacerbated by the fact that current LLMs such as ChatGPT are owned by private companies.  This raises significant data ownership issues---the terms of use of many of these tools transfer ownership of any input into their systems to these companies. Providing feedback to students using LLMs inherently requires giving these companies a copy of that assignment and the rights to use that data as they see fit---most likely in a scenario in which students are unable to opt out of the provision of this feedback.  It is this data transfer that has already led to organizations forbidding the use of generative AI tools in their work---for instance the Australian Research Council does not allow the use of these tools in their grant review processes~\cite{ARC2023}.

While generative AI models may be intended for universal use, they are not always universal in their accessibility.  Just because modern university students are digital natives does not mean that they are digitally fluent; interacting with automated feedback systems will be another skill that needs to be developed.

Access to LLM-based chatbots requires reliable access to the internet, as well as whatever subscription arrangements are required for the tools themselves.  While these are near-universal in some contexts, these can serve as significant barriers in poorer areas, or for students in less developed economies. Equitable access to online tools will also require all of the considerations of accessible technology.  Accessibility considerations are particularly important in the context of automated feedback---the students most likely to benefit from the advantages of automated feedback such as asynchronicity and infinite patience are those most likely to be in the long tail of ordinary feedback requirements.

\subsection{Development}
Automated feedback is only as good as the models used to provide it, and the models are only as good as the data that are used to train them. LLMs provides a specific challenge in this regard, in that it is trained on a wide range of available data, roughly made up of trillions of words available on the internet, e.g., Gemma 2~\cite{gemma2}, rather than on a curated sample of similar assignments, with corresponding expert feedback.  It is a generalized model, not a specifically tailored model, and as such it runs the risk of providing feedback that may be appropriate in a generic context but not in the specific discipline context.

A solution will be the development of automated feedback models that are specifically trained to provide such feedback, as outlined in the previous section.  These will have the advantage that they are purpose built and can therefore be developed to provide appropriate feedback; but the appropriateness will be dependent upon how they are trained. Furthermore, even with such a tailored development procedure, generative LMs have a core technological issue in terms of hallucination, implying the absence of a guarantee of factuality. There are furthermore well-known examples in which bias in training datasets have led to issue with classification, such as facial recognition systems that have been trained on predominantly white datasets.  It will be critically important to develop appropriate training datasets if automated feedback is to provide fair and equitable value to all students.

Following the development pipeline outlined in the previous section, the training procedure requires two datasets---one containing samples of student work and the learning curriculum, and one containing samples of the appropriate feedback to provide to the students, provided in RLHF. Training sets for student assessment are usually convenience samples of historical data---the assignments of current and previous students are used to train the models to provide feedback on future submissions.  While these make excellent test data, they are unfortunately not ideal as training data. Student assignments are not evenly distributed with regard to their needs for feedback.  The most common errors are by their definition common, and therefore easier to detect and to automate.  The least common errors, which perhaps are those in most need of expert feedback, are the least common and thus the hardest to capture and automate in the model.  Care must be taken to ensure that the models that are developed are able to provide useful feedback even when uncommon, or even unfamiliar submissions are provided---rather than just indulging in model hallucinations. An algorithm that is most accurate on average may systematically discriminate against a specific minority~\cite{johri2023ethical}.

The second dimension of a comprehensive training set is the provision of sample feedback from experts. Accessing expertise raises questions such as who is an expert, and who gets to decide who an expert is.  There are also operational questions such as how this expertise should be captured---whether as a convenience sample from historical submissions or through a deliberate process to solicit expert feedback for the purposes of building the model.

The key challenge is to capture sufficient expertise so that the whole range of possible assignment submissions can be addressed---to ensure coverage over the whole spectrum of student submissions, without risking the loss of the long tail.  Expertise is valuable and thus expensive, leading to a natural tendency to focus it on the most highly leveraged part of the spectrum; but to do so is to risk developing automated feedback systems that are just memorizing rather than synthesizing. Due to this expense, it will also be key to employ transfer learning methodologies so as to take advantage of pre-existing developments. Since similar courses will potentially overlap in terms of the feedback needed, there will be possibilities for significant operational savings by basing a model for a new computer engineering degree on one developed for a similar engineering degree program.

\subsection{Impact of Automated Feedback on Learning}
It is critical to also consider the impact that automated feedback will have upon student learning.  Generative AI provides the opportunity to provide repeatable, scalable and instant automatically generated feedback to students, making abundant a previously scarce and expensive learning resource.  When coming from a place of scarcity, the prospect of abundance is inherently attractive; but the longer term consequences must also be considered, along with who actually benefits from the abundance.

Feedback on assignments is a critically valuable part of the learning process and the timelier the feedback, the more valuable.  For the majority of students, and the majority of feedback they require, generative AI tools represent an opportunity to make that feedback instant. In the early implementations, this represents an opportunity to improve student learning. However, as they become accustomed to on-demand instant feedback, it is likely that students’ attitudes towards feedback will change. A core challenge here is whether this abundance of feedback will devalue it, and whether automatic feedback will be seen as equivalent to human feedback. If there appears to be an inexhaustible supply of feedback, how much will students consume?  Part of the value of feedback is that it triggers the reflective cycle in students to improve their work; if they can instead simply make some changes, resubmit, and get new feedback, will they still reflect deeply on their work? There is therefore a risk that abundant feedback could lead to a deskilling of our students.  Rather than develop their own reflective competences, and the ability to review their own work, they may become reliant upon having external tools to guide the development of their work---in effect gamifying their work into pleasing the generative AI tool rather than developing their actual skills.

These quicker feedback loops also introduce a risk of homogenizing towards good practice.  Generative AI systems are valued because they are able to quickly amplify the performance of workers, and they have been shown to be of greater benefit to novices than to experts~\cite{mollick2024co}.  These quicker gains are because the systems push users towards a standard of good practice that has emerged from the training set.  While this is helpful in speeding the development of novices, it risks pushing all users towards that particular version of good practice, regardless of what feedback they actually require.

It is not yet clear whether feedback is fungible.  Do students regard feedback from a generative AI the same way they regard feedback from a human?  Is AI-generated feedback less ``real''?  Will students have the same expectations of AI-generated feedback?  Society holds self-driving cars to much higher standards than they do human drivers, but we do not yet know if students will similarly hold different standards for AI-generated feedback.

Ultimately it is essential that students trust the feedback they are receiving.  Social media (an incredibly abundant resource) have shown that people are willing to dismiss information that does not fit with what they already believe.  Will students be less likely to accept feedback that does not reinforce what they already believe if it comes from an AI-generated source?  This is the kind of feedback that they most need to trust; but will they trust it if it does not come from a human?

Establishing trust will be an essential part of rolling out a generative AI feedback system.  Universities have systems that implicitly establish trust in traditional feedback models---feedback comes from a human who has been hired into the system for their expertise in the area in which they are providing feedback.   Trust will depend upon whether the generative AI system has been customized for the specific application.  For LLMs, it will depend on whether the feedback is sufficiently specific to be valuable, despite having been trained on generic datasets.  For specialized tools, it will depend upon whether the training process is sufficiently transparent for users to understand how the feedback has been generated. Trust will also depend on whether the generative AI provides feedback that is helpful.  A system that provides only hallucinations will quickly be dismissed; but a system that has been shown to be able to help students learn will be trusted to continue to do so. This further enforces the need to develop language models which can guarantee the factuality and relevance of their outputs.

Trusting the feedback will be easier if you are the median student, in need of the median feedback; but for those students in the long tail there is a risk that generative AI systems exacerbate the inequalities already in our systems. Human generated feedback is able to respond to the full range of student assignment submissions.  While the long tail may be less common, it is not inherently problematic for humans to provide meaningful feedback.

For generative AI systems, however, uncommon submissions pose a problem.  What feedback should a generative AI system give to an assignment that is unlike any other it has been trained upon?  This is a significant risk given that many specialized systems are trained on historical data, which are a biased sample of potential student submissions.

Students in the long tail are the ones needing the most help; theirs is the practice that is the furthest (either behind or ahead of) from best practice, and the ones who need customized feedback the most.  There is a risk that generative AI systems are unable to provide meaningful feedback to such students, or worse, to provide misleading feedback instead.  These risks can be mitigated through more comprehensive training datasets; but this represents an additional upfront cost that will benefit smaller and smaller numbers of students.

\subsection{Evolution over Time}
It is important to consider how automated feedback systems can and will be used over time.  While it is perhaps impossible to correctly predict how circumstances will change over time, it is nonetheless irresponsibly naïve to assume that they will not change at all. Changing circumstances are often used to justify a kind of techno-optimist---these tools are here, they will be used anyway, and so we should embrace them.  There are grounds to believe this~\cite{Shaw2023} but this does not allow us to ignore questions about how widely they should be used, and for which students.  Optimism cannot simply be a naïve alternative to the ``slippery slope'' argument.

Drift over time is a challenge for generative AI models.  These models will require maintenance to ensure that their feedback remains relevant given temporal changes in context, theory, and prior learning profiles of student cohorts.  This maintenance will potentially be expensive, and there is a risk that institutions will not invest in this maintenance.

This risks the possibility that automated feedback engines will in fact calcify the learning experience---having invested heavily in building a feedback engine, they will be reluctant to update their teaching materials to move away from the set point where they can provide the detailed feedback.

A key theme for all AI decision making systems is the extent to which they operate as decision support systems rather than decision making systems.  Will there be humans in the loop, and to what extent will these humans retain control of the process? This decision should be dependent upon the jeopardy involved---what are the consequences of a wrong decision? Are we more afraid of a false positive than a false negative?  The provision of automated feedback to students carries potentially less jeopardy than the automation of assessment outcomes.  A wrong grade on an assignment has greater consequence that inaccurate feedback, particularly when the alternative to automated feedback may be no feedback at all.

Initially, the distinction between marking and feedback is clear; however there is a risk that the lines between the two blur over time.  Some types of feedback will correlate strongly with high performance; other types of feedback will correlate with failure.  Over time, if this correlate stays sufficiently strong, the automated feedback engine drift into become an automated assessment engine---without the human in the loop.

This potential drift is a real risk over time.  The more accurate a decision support system becomes, the more likely it is to morph into a decision making system---for how long will an institution continue to carry the expense of checking the recommended decisions when so few of them are actually incorrect?  How much will they be willing to pay to avoid each false decision?  But which false decisions will they be avoiding, and will they be in the long tail?

Resource reallocation is often presented as a driver for the automation of feedback.  After an upfront investment we can move from a model of scarce, expensive feedback into a paradigm of abundant feedback, and instead reallocate our resources to supporting our students in different ways.  But do we believe the reallocation narrative?  We all say that this is an opportunity to redirect resources to more valuable learning situations, but is that likely to actually happen?  Or will it just fund more postdocs elsewhere instead?  While there is immense potential for more efficient resource allocation, we must ask ourselves if we can still uphold our graduate standards if we solely focus on automatable outcomes~\cite{johri2023ethical}.

Many of these ethical concerns actually represent tensions between two good outcomes that we want to achieve, such as the balance of more complex models potentially giving better accuracy at the cost of not being able to understand the models as easily.  How we balance competing outcomes is a choice, and it can no longer just be a default choice---it needs to be a deliberate choice, with explicit frameworks to guide that thinking.

\subsection{Scaling Up}
Maintaining ethical oversight in AI systems is a challenging problem, especially as they scale-up \cite{kodakandla2024scaling}. Scaling up can lead to lack of insights into who is using the system and for what purpose. Scaling is also problematic because a small problem within the system can have a dramatically high impact as a larger number of users are affected \cite{barmer2021scalable}. To ethically scale-up assessment systems, it is important to first and foremost test any application before implementing it. Rigorous testing and evaluation are the cornerstone of quality software systems, and AI-based assessment applications are no exception. In this case, though, given the challenges with bias and fairness, testing and evaluation should involve a diverse set of users. Providing good documentation and ensuring transparency and explainability can further support this process by building trust among users and allowing them to understand how decisions are made and have confidence in reporting any misgivings and problems as soon as they occur. This can also help build guidance and guardrails around the acceptable use of an application \cite{mcdonald2025generative}. It is important then to conduct regular audits of the systems in addition to a process to continuously monitor the application. If needed, third-party audits can also be included in the process. Finally, it is important to remember throughout this process that the application, as well as its ethical oversight, should complement rather than replace human educators. Given the goal of education to support learning by fostering critical thinking and advancing the personal development of learners, it is important that educators are part of the process throughout. 

\subsection{AI Literacy}
Finally, one of the critical assumptions made here is that not only those who are building AI-based assessment applications, but even those who are using it, will be AI literate. Given the fast-changing AI environment, this is an assumption as best, and to ensure the best use of AI systems, it is important that educators are AI literate. As the use of generative AI applications has indicated, educators are often far behind students in the adoption of AI and often lack a fundamental understanding of how AI works \cite{nguyen2025use}. Without this knowledge, interventions based on AI can never be fully adopted within educational settings. Therefore, training for educators is essential for proper use of AI systems, especially in an ethical and responsible manner. Faculty need to understand both the technical aspects of how a system work, to some extent, and how the overall application fits within their ecosystem \cite{schleiss2024integrating}. And even if they do not develop and understanding of how to create such a application, knowing how it works is critical in assessing its output and usefulness for the task at hand \cite{almatrafi2024systematic}.

\section{Conclusion}
Recent advancements in generative AI, including tools like ChatGPT, offer transformative possibilities in higher education by potentially turning feedback from a scarce, expensive resource into an abundant one. While this paradigm shift presents enormous potential for enhancing learning, it also risks uneven distribution of benefits.

Developing generative AI tools requires significant resources and may introduce biases dependent on the datasets that are used to train them with. These biases can affect both the overall suitability of AI for teaching and its ability to address individual student needs. The advantage of automation in scaling expertise is counterbalanced by the risk of prioritizing easily automatable aspects, potentially neglecting students with unique needs and entrenching existing educational disadvantages.

To navigate these challenges, this paper draws inspiration from the RESPACT framework for potentially navigating these challenges as faculty implement generative AI tools.  This framework provides a lens for considering the identified ethical issues (and others that may emerge), considering them along the dimensions of the RESPACT framework. A deliberate framework can provide the structure needed to make good decisions, and in doing so provide the faith in the tools to support better learning for your students. 

In conclusion, this paper outlines four critical ethical considerations in implementing generative AI for automated feedback, recognizing that many issues involve trade-offs. For instance, more sophisticated models may provide better feedback but at the cost of transparency and explainability. We encourage future work to build upon these ethical considerations, ensuring responsible development of automated feedback methodologies based on generative language modeling, ultimately enabling all students to benefit from the opportunities these models provide.


%

\appendices


\section*{Acknowledgments}
EDL, MZ, and JB are supported by the research grant (VIL57392) from VILLUM FONDEN. AJ is partly supported by US NSF Awards 2319137, 1954556, and USDA/NIFA Award 2021-67021-35329. Any opinions, findings, and conclusions or recommendations expressed in this material are those of the authors and do not necessarily reflect the views of the funding agencies.

\ifCLASSOPTIONcaptionsoff
  \newpage
\fi



\bibliographystyle{IEEEtran}
\bibliography{bibtex/bib/custom}
%

%








\end{document}